\documentstyle[12pt]{article}
\begin{document}
\def\beq{\begin{equation}}
\def\eeq{\end{equation}}
\def\beqa{\begin{eqnarray}}
\def\eeqa{\end{eqnarray}}
\def\gapp{\stackrel{>}{~}}
\begin{flushright}
FTUV/97-12\\
IFIC/97-12\\
PRL-TH-97/7 \\
hep-ph/9703380 
\end{flushright}
\vskip .25cm
\begin{center}
{\large \bf \boldmath
Forward-backward asymmetry in $e^+e^- \rightarrow \nu\overline{\nu}Z$
from anomalous CP-odd $WWZ$ couplings
}
\vskip .5cm
Saurabh D. Rindani \footnote{
Permanent address: {\it Theory Group, Physical Research Laboratory,
Nav\-rang\-pura, Ahmed\-abad 380 009, India.} 
E-mail address: {\tt saurabh@prl.ernet.in}}\\
{\it Instituto de F\' \i sica Corpuscular - CSIC \\
Departament de F\' \i sica Te\` orica, Universitat de Val\` encia \\
46100 Burjassot (Val\` encia), Spain}
\vskip .25cm
and \\
\vskip .25cm
J.P. Singh\\
{\it Physics Department, Faculty of Science \\
M. S. University of Baroda, Vadodara 390 002, India \\
}
\vskip .5cm
{\bf Abstract}
\end{center}
\vskip .25cm

Anomalous CP-odd $W^+W^-Z$ couplings can give rise to a for\-ward-back\-ward
asymmetry in the $Z$ angular distribution in the $e^+e^-$ centre-of-mass frame 
in the process $e^+e^- \rightarrow \nu\overline{\nu}Z$. Of the three CP-odd 
couplings possible, only the imaginary part of one of the couplings, $f_4^Z$,
which is C odd and P even, contributes to the forward-backward asymmetry. It 
is found that a limit of the order of 0.1 can be placed on this
coupling at a Next Linear Collider with centre-of-mass energy of 500 GeV 
and integrated luminosity of 50 fb$^{-1}$.

\vskip .5cm
\vskip .2cm
\newpage

The second phase of the Large Electron Positron collider (LEP) has begun 
operation. One of the tasks that this LEP2 phase will undertake is the 
investigation of anomalous $W^+W^-\gamma$ and $W^+W^-Z$ couplings through the
process $e^+e^- \rightarrow W^+W^-$ \cite{LEP2}. This investigation 
would be able to
improve limits on such anomalous couplings which might arise in scenarios beyond
the standard model (SM). There have also been suggestions to study CP-violating
vector boson couplings in $e^+e^- \rightarrow W^+W^-$ at LEP2 \cite{Chang}.

While $e^+e^- \rightarrow W^+W^-$ can test only a combination of $W^+W^-\gamma$
and $W^+W^-Z$ couplings, the process $e^+e^- \rightarrow \nu\overline{\nu}Z$
would be better suited to probe the $WWZ$ couplings separately. Though this
latter process is not feasible of observation at LEP2, it would be significant
at higher energies ($\sqrt{s} \gapp 300$ GeV) of  a future Next Linear Collider
(NLC). Studies exist on the sensitivity of NLC to anomalous $WWZ$ couplings 
through the reaction $e^+e^- \rightarrow \nu\overline{\nu} Z$ \cite{Bori,
Ambro,deb}.
 However, these
are restricted only to the CP-conserving couplings. We investigate in this work
a simple CP-violating asymmetry, viz., the forward-backward (FB) asymmetry in 
the angular distribution of the $Z$ in the $e^+e^-$ centre-of-mass (cm) frame,
as a signal for CP-odd $WWZ$ couplings.

In general, the CP-violating part of the most general Lorentz-invar\-iant 
\mbox{effective} $WWZ$ coupling $i\,g_{WWZ}\,\Gamma^{\alpha\beta\mu}_Z 
(q,\overline{q},p)$, re\-pre\-sent\-ing the pro\-cess 
$Z_{\mu}(p)\\ \rightarrow W^-_
{\alpha} (q) + W^+_{\beta} (\overline{q})$,  is given by \cite{Hagi}
\beq
\Gamma^{\alpha\beta\mu}_{Z,{\rm CP-odd}} (q,\overline{q},p) = 
i f^Z_4 (p^{\alpha} g^{\mu\beta} + p^{\beta} g^{\mu\alpha} ) - f^Z_6 
\epsilon^{\mu\alpha\beta\rho} p_{\rho} - \frac{f^Z_7}{m^2_W} (q-\overline{q})
^{\mu} \epsilon^{\alpha\beta\rho\sigma} p_{\rho} (q-\overline{q})_{\sigma}
\eeq
The couplings $f_4$,$f_6$ and $f_7$ (we suppress the superscript $Z$ now 
onwards) 
are momentum dependent and complex in the general case. The overall coupling 
constant $g_{WWZ} = - g \cos\theta_W$, where g is the usual semi-weak coupling,
and $\theta_W$ is the weak mixing angle.

The forward-backward (FB) asymmetry of $Z$, related to an asymmetry in 
the variable
\beq
\cos \theta_Z = \frac{(\vec{p}_{e^+} - \vec{p}_{e^-})\cdot \vec{p}_Z} {2 \vert \vec{p}_{e^-}\vert
\vert \vec{p}_{Z}\vert }
\eeq
in the cm frame, is odd under CP. This follows from the fact that under C,
$\vec{p}_{e^-} \leftrightarrow \vec{p}_{e^+} $, $\vec{p}_Z \rightarrow \vec
{p}_Z$, and under P, $\vec{p}_{e^-} \rightarrow - \vec{p}_{e^-}$, $\vec{p}_{e^+}
\rightarrow - \vec{p}_{e^+}$, $\vec{p}_Z \rightarrow  -\vec{p}_Z $. However, 
$\cos \theta_Z$ is even under naive time-reversal ${\rm T}_{\rm N}$, which, 
like P, reverses
all momenta. Unlike genuine time-reversal, however, ${\rm T}_{\rm N}$ does not interchange the 
initial and final states. $\cos \theta_Z$ is therefore 
${\rm CPT}_{\rm N}$ odd. 

Let us see what the CPT theorem implies for the dependence of an asymmetry in
the variable $\cos \theta_Z$ on the parameters of the effective Lagrangian.
Observation of a ${\rm CPT}_{\rm N}$-odd quantity like $\cos \theta_Z$ does
not necessarily imply conflict with the $CPT$ theorem since there is no
interchange of initial and final states involved. 
However, in order to avoid conflict with the CPT theorem, quantities 
which are ${\rm CPT}_{\rm N}$-odd can have nonzero values only if the amplitude
has an absorptive part. In its absence, the transition matrix is effectively
Hermitian, and an interchange of initial and final states envisaged in genuine
time reversal does not change
the amplitude except for an overall phase factor. The transition amplitudes then
have the same essential behaviour under
T and T$_{}$N. In an effective Lagrangian approach, where we calculate
amplitudes only at tree level, this absorptive part could only arise if the 
couplings in the effective Lagrangian are complex. It can be seen from the 
effective couplings in eq. (1) that the absorptive contribution could come from
Im$f_4$, Re$f_6$ and Re$f_7$, since without absorptive parts, $f_4$ would
be real, and $f_6$, $f_7$ purely imaginary from Hermiticity.
Thus, from considerations of the CPT theorem, an asymmetry in $\cos\theta_Z$ 
can in principle 
get nonzero contributions only from Im$f_4$, Re$f_6$, and Re$f_7$. 

We find, 
however, that if the azimuthal angle of $Z$ is integrated over, as in the 
FB asymmetry, only Im$f_4$
contributes. A search for the FB asymmetry will therefore
enable limits to be placed on the single parameter Im$f_4$.\footnote{For a 
suggestion of an experiment to isolate the couplings $f_6^{\gamma}$ and 
$f_6^Z$, see \cite{spanos}. It should be noted that an analogous FB asymmetry
in $e^+e^- \rightarrow \nu\overline{\nu}\gamma$ is absent, because an 
$f_4^{\gamma}$ term in the $W^+W^-\gamma$ couplings is forbidden because of 
electromagnetic gauge invariance.
}

We have calculated the squared matrix element for the process
\beq
e^- (p_1) + e^+  (p_2) \rightarrow \nu (k_1) + \overline{\nu} (k_2) + Z (q)
\label{reaction}
\eeq
in the standard model, together with the anomalous CP-odd couplings Im$f_4$,
Re$f_6$ and Re$f_7$, keeping only the linear terms in the anomalous couplings.
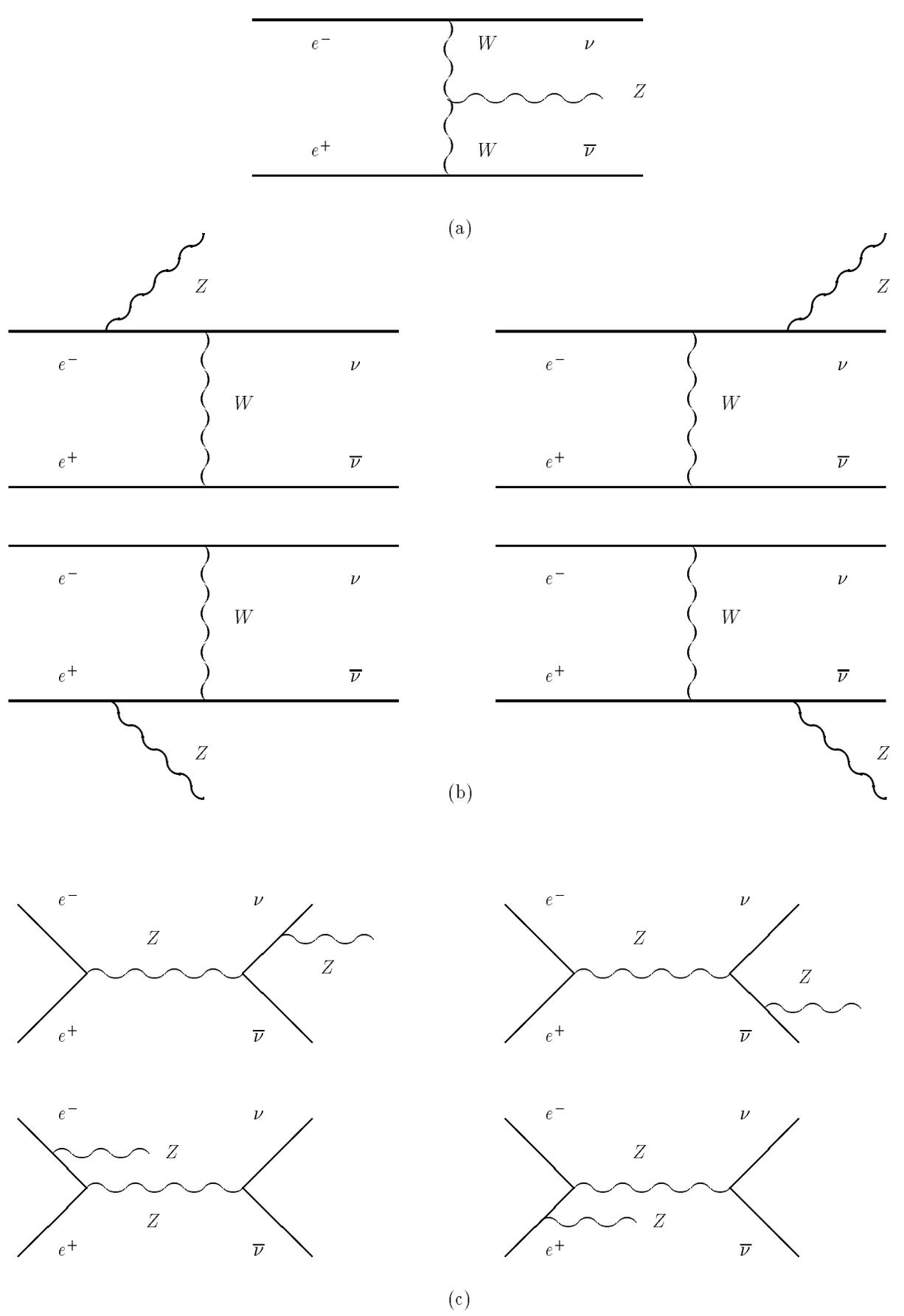
The Feynman diagrams contributing to the process (\ref{reaction}) are shown 
in Fig. 1. 
We drop contributions from the SM diagrams which involve $Z$ exchange
in the $s$ channel (Fig. 1 (c)). 
These contributions in 
SM are known to be negligible for $\sqrt{s}>300$ GeV \cite{Ambro}. 
To ensure that this 
remains true in our case, we can also impose a cut on the $Z$ energy which 
ensures that the $\nu\overline{\nu}$ invariant mass is sufficiently far from 
the $Z$ mass \cite{deb}. We choose to apply a somewhat simpler cut, however.
We apply 
a cut of $\vert E_Z -\sqrt{s}/2\vert > 5\Gamma_Z$, and 
have checked that for $\sqrt{s}>300$ GeV, this cut
keeps the SM contribution to the cross section from
the virtual $Z$ diagrams to less than 1\%.

The phase space-integral is carried numerically using Monte Carlo integration. 
It is found that the only nonzero contribution to the forward-backward 
asymmetry comes from Im$f_4$. Re$f_4$ as well as the other 
CP-violating couplings give
vanishing asymmetry because of the integration over the azimuthal angles.

We present below the squared matrix element for (\ref{reaction}) in the 
approximation mentioned above. The expression for the SM contribution is rather
long, and we refer the reader to ref.\cite{Ambro}, where full expressions may be
found. Our results agree with those of \cite{Ambro}. The interference between 
the matrix elements of the $WWZ$ diagram (that in Fig. 1(a)) from SM 
and the same diagram from 
anomalous CP-violating couplings is (we present only the part proportional to
Im$f_4$):
\beqa
\lefteqn{ 2 \overline{\sum}{\rm Re}\left( M_{\rm SM}^{WWZ}M_{\rm anom}^{WWZ*}
\right) 
 = \frac{{\rm Im}f_4\,\, g^6\, \cos^2\theta_W}{32\,(2p_1.p_1' + m_W^2)^2(2p_2.p_2'+m_W^2)^2}
\left[ (p_1-p_1') . k  \right. } \nonumber \\
& \times &\left. \biggl\{ \frac{8}{m_Z^2}\left(
  p_1' . p_2'
p_1 . k p_2 . k + p_1 . p_2 p_1' . k p_2' . k 
-p_2 . p_2' p_1 . k p_1' . k 
\right. \right.   \nonumber\\
&& \left. \left.
- p_1 . p_1' p_2 . k p_2'
. k \right) + 4 \left( p_1 . p_1' p_2 . p_2' + p_1' . p_2 p_1 . p_2'
- p_1 . p_2 p_1' . p_2' \right)\biggr\} \right. \nonumber\\
& & \left.-  4 p_1 . p_1' p_2 . k p_2' . k +12 p_2 . p_2' p_1 
. k p_1' . k  - 4 \left( p_1'.p_2' p_1.k p_2.k + p_1.p_2 p_1'.k p_2'.k \right)
\right.  \nonumber \\
&& \left. 
- 2 m_Z^2 \left( p_1.p_1' p_2.p_2' + p_1.p_2' p_1'.p_2 - p_1.p_2 p_1'.p_2'
\right) \right]
\eeqa

The inter\-ference of the matrix element from the SM involving $Z$ 
brems\-strahlung
from $e^{\pm}$, $\nu$, $\overline{\nu}$ lines (diagrams in Fig. 1(b) )
and the anomalous $WWZ$ diagram (that of Fig. 1(a))
is:
\beqa
\lefteqn{
2 \overline{\sum}{\rm Re}\left( M_{SM}^{Zbrem}M_{anom}^{WWZ*}\right)}
\nonumber \\
& = &
\frac{ {\rm Im}f_4\,\, g^6}{64(2p_1.p_1' + m_W^2)(2p_2.p_2' + m_W^2)}
 \Biggl[ 
(-1+2\sin^2\theta_W )  \nonumber \\
&  \times &\left\{
\frac{T_1}{(m_Z^2-2p_1.k)(2p_2.p_2'+m_W^2)}
+ \frac{T_2}{(m_Z^2-2p_2.k)(2 p_1.p_1' + m_W^2)} \right\} \nonumber \\
 &  + &\left\{ \frac{T_3}{(m_Z^2+2p_1'.k)(2 p_2.p_2' + m_W^2)} 
+ \frac{T_4}{(m_Z^2+2p_2'.k)(2 p_1.p_1' + m_W^2)} \right\} \Biggr] ,
\eeqa
where $T_i$ are given by 
\beqa
T_1 & = & 4 \left[ ( p_1.p_2' - p_2'.k) ( p_1.p_1' p_2.k - p_1.p_2 p_1'.k) -
	p_1'.p_2' p_1.k (p_2.k + p_1.p_2) \right. \nonumber \\
&&\left.
	+ p_1.k p_2.p_2' (p_1'.k + p_1.p_1')
	\right] +\frac{8}{{m_Z^2}}\left[
	(p_1.k)^2 (p_1'.p_2' p_2.k - p_1'.k p_2.p_2') \right. \nonumber \\
&& \left.  + p_1.k p_2'.k ( p_1.p_2
	p_1'.k -p_1.p_1' p_2.k) \right],
\eeqa
\beqa
T_2 & = & 4 \left[-( p_1'.p_2 - p_1'.k) ( p_2.p_2' p_1.k - p_1.p_2 p_2'.k) +
	p_1'.p_2' p_2.k (p_1.k + p_1.p_2) \right. \nonumber \\
&&\left.
	- p_2.k p_1.p_1' (p_2'.k + p_2.p_2')
	\right] -\frac{8}{{m_Z^2}}\left[
	(p_2.k)^2 (p_1'.p_2' p_1.k - p_2'.k p_1.p_1') \right. \nonumber \\
&& \left.  + p_1'.k p_2.k ( p_1.p_2
	p_2'.k -p_2.p_2' p_1.k) \right],
\eeqa
\beqa
T_3 & = & 4 \left[( p_1'.p_2 + p_2.k) ( p_1.p_1' p_2'.k - p_1'.p_2' p_1.k) +
	p_1.p_2 p_1'.k (p_2'.k - p_1'.p_2') \right. \nonumber \\
&&\left.
	- p_1'.k p_2.p_2' (p_1.k - p_1.p_1')
	\right] +\frac{8}{{m_Z^2}}\left[
	(p_1'.k)^2 (p_1.p_2 p_2'.k - p_1.k p_2.p_2') \right. \nonumber \\
&& \left.  + p_1'.k p_2.k ( p_1'.p_2'
	p_1.k -p_1.p_1' p_2'.k) \right],
\eeqa
\beqa
T_4 & = & 4 \left[- ( p_1.p_2' + p_1.k) ( p_2.p_2' p_1'.k - p_1'.p_2' p_2.k) -
	p_1.p_2 p_2'.k (p_1'.k - p_1'.p_2') \right. \nonumber \\
&&\left.
	+ p_2'.k p_1.p_1' (p_2.k - p_2.p_2')
	\right] -\frac{8}{{m_Z^2}}\left[
	(p_2'.k)^2 (p_1.p_2 p_1'.k - p_2.k p_1.p_1') \right. \nonumber \\
&& \left.  + p_1.k p_2'.k ( p_1'.p_2'
	p_2.k -p_2.p_2' p_1'.k) \right].
\eeqa
The differential cross section for the process to first order in Im$f_4$ is
given by
\beqa
\frac{d\sigma}{d\cos\theta_Z} 
& = &\frac{1}{2\pi}\int \overline{\sum}
\left[ \vert M_{SM}^{WWZ} + M_{SM}^{Zbrem} \vert ^2 \right. \nonumber \\
&&\left. +2 {\rm Re}\left( (M_{SM}^{WWZ} + M_{SM}^{Zbrem}) M^{WWZ*}_{anom}
\right)
\right] dE_Z d\cos\tilde{\theta}d\phi,
\eeqa
where $E_Z$, $\theta_Z$ and $\phi$ are respectively the $Z$ energy, polar and
azimuthal angles in the c.m. frame, and $\tilde{\theta}$ is the polar angle
of the neutrino in the $Z$ rest frame.

The forward-backward asymmetry of the $Z$ with respect to the beam direction,
with $\theta_Z$ integrated over the ranges $\theta_Z > \theta_0$ and $\theta_Z < \pi - \theta_0$, is defined by 
\beq
A_{FB}(\theta_0) = \frac{1}{\sigma (\theta_0)} \left[ \int^{\cos\theta_0}_0
\frac{d\sigma}{d\cos\theta_Z}\,d\cos\theta_Z
-  \int_{-\cos\theta_0}^0
\frac{d\sigma}{d\cos\theta_Z}\,d\cos\theta_Z
\right],
\eeq
where
\beq
\sigma (\theta_0) = 
\int_{-\cos\theta_0}^{\cos\theta_0}
\frac{d\sigma}{d\cos\theta_Z}\,d\cos\theta_Z.
\eeq
As discussed before, $A_{FB} (\theta_0)$ is proportional to Im$f_4$. A cut on
$\theta_Z$ is imposed because in practice, it would be difficult to observe a
$Z$ near the forward or backward directions.

We have evaluated $\sigma (\theta_0)$ and $A_{FB} (\theta_0)$ for different 
energies and $\theta_0=20^{\circ}$. Fig. 2 shows $\sigma (\theta_0
)$ as a function of $\sqrt{s}$. In Fig. 3, we have plotted $A_{FB} (\theta_0)/
{\rm Im}f_4$ as a function of $\sqrt{s}$.

We find that the asymmetry is of the order of $10^{-1}$ in units of Im$f_4$.
This information can be converted into a 95\% confidence level (CL) limit 
on Im$f_4$
that can be placed if an asymmetry is not seen, by demanding that in the limiting
case, the difference in the number of forward and backward events is 1.96  times
the statistical error in the total number of events:
\beq
L \sigma (\theta_0) A_{FB} (\theta_0) = 1.96  \sqrt{L \sigma (\theta_0)},
\label{95cl}
\eeq
where $L$ is the integrated luminosity. Writing 
\beq
A_{FB} (\theta_0) = {\rm Im}f_4 \cdot C(\theta_0),
\eeq
eq.(\ref{95cl}) leads to the 95\% CL limit on Im$f_4$ of
\beq
{\rm Im}f_4 <  \frac{1.96}{C(\theta_0) \sqrt{L \sigma (\theta_0)}}.
\eeq
Table 1 shows the 95\% CL limit that could be achieved on Im$f_4$ 
with integrated luminosities of 10 fb$^{-1}$, 50 fb$^{-1}$ and 100 fb$^{-1}$
for different energies, and an angular cut  of $\theta_0 = 20^{\circ}$.

\begin{table}
\begin{center}
\begin{tabular}{||c|c|c|c|c||} 
\hline
 $\sqrt{s}$  & Cross section &  \multicolumn{3}{c||}{  95\% CL limits}\\
(GeV)          &   (fb) &(a) & (b) & (c)\\
\hline
 300 &    48.0   &    .40  & .18 &   .13\\
 400 &    114   &    .27  & .12 &   .087\\
 500 &    186   &    .24  & .11 &   .075\\
 600 &    256   &    .23  & .10 &   .072\\
 700 &    323   &    .22  & .10 &   .071\\
 800 &    385   &    .23   & .10 &   .071\\
 900 &    441   &    .23   & .10  &   .073\\
1000 &    494   &    .24 & .11 &   .074\\
1500 &    692  &    .27 & .12 &   .087\\
2000 &    824  &    .32 & .14 &   .10\\
\hline
\end{tabular}
\end{center}
\caption{Cross sections and 95\% CL limits on Im$f_4$ which could be 
obtained for integrated 
luminosities (a) 10 fb$^{-1}$, (b) 50 fb$^{-1}$ and (c) 100 fb$^{-1}$. An 
angular cut-off of 20$^\circ$ has been imposed on the forward and backward
Z directions.}
\end{table}

We thus see that it is possible to achieve a sensitivity of 0.24
to 7.5$\times 10^{-2}$  in
Im$f_4$ for an NLC operating at $\sqrt{s}=500$ GeV with luminosity in the 
range 10 to 100 fb$^{-1}$. For somewhat higher $\sqrt{s}$, the 
sensitivity is
improved further, though not drastically. For very large $\sqrt{s}$ ($\sqrt{s}
\approx 2000$ GeV), the sensitivity starts to decrease. 

We conclude with a few remarks.

We have studied a CP-violating asymmetry which depends on the imaginary part
of a single effective coupling, namely, $f_4$. While our approach has been to
purely phenomenological, where we do not assume any definite model, it would
be worthwhile to give a little thought to possible scenarios which could
give rise to non-negligible Im$f_4$. In gauge models, such anomalous
couplings are expected to arise only at the loop level, and therefore expected
to be extremely small. Possible contributions to $f_4$ have been investigated
in a general scenario in \cite{burgess}. While their investigation is for 
LEP 200 energies, and restricted to the dispersive contributions below
new thresholds, we might extrapolate their reasoning to conclude that 
there could be large scalar or fermion-loop contributions at $\sqrt{s}=500$ GeV 
provided there are new particle thresholds in that region. These contributions
can arise in models with exotic fermions or extra Higgs with masses in the 
region of a few hundred GeV, which are not ruled 
by present experiments. Even then, the 
numerical value of Im$f_4$ is still expected to be at least an order of 
magnitude smaller than the limit we quote as feasible. If the signal we discuss
is discovered, it would indeed signal very new physics. This situation perhaps
calls for a search for more sensitive tests for studying CP-violating couplings
like $f_4$.

We have assumed ideal conditions, viz., 100\% efficiency for energy and angle
determination of $Z$. While a leptonic decay channel of $Z$ can lead to 
an accurate determination of the energy and angle, there would be some loss 
of efficiency because of angular cuts in the forward and backward directions
on the leptons. Effects of initial-state radiation and off-shell $Z$ effects
can also lead to some amount of deterioration in the sensitivity. These
effects and the effects of cuts to remove the SM background need further study.
However, we expect the quantitative conclusions to remain valid to a good 
deal of accuracy.

We have thus demonstrated the possibility of measuring separately 
(the imaginary part of) the CP-odd coupling $f_4$ at NLC by means of a simple
forward-backward asymmetry.
\vskip .2cm

\noindent {\bf Acknowledgements} One of the authors (S.D.R.) thanks Prof. 
J.W.F. Valle
for his warm hospitality at the University of Valencia where most of this work
was done. He also thanks Debajyoti Choudhury for discussions. 
The other author (J.P.S.) thanks 
colleagues and authorities of Jiwaji University, Gwalior, where most of 
the work was done. 
The authors also thank Prof. R. Rajaraman for seeding an initial collaboration 
which evolved into this work.
The work of S.D.R. has been supported in part by DGICYT under 
Grant Ns. PB95-1077 and SAB95-0175, as well as by the TMR network 
ERBFMRXCT960090 of the European Union. 
J.P.S. acknowledges financial assistance from D.S.T.,
New Delhi, during the first phase of this work.

\end{document}